\documentclass{article}

\usepackage[preprint]{neurips_2023}
\usepackage{
  amsmath,
  amsthm,
  amssymb,
}
\usepackage{graphicx}
\usepackage{subcaption}
\usepackage{float}
\usepackage{placeins}

\usepackage[utf8]{inputenc} 
\usepackage[T1]{fontenc}    
\usepackage{url}            
\usepackage{booktabs}       
\usepackage{amsfonts}       
\usepackage{nicefrac}       
\usepackage{microtype}      
\usepackage{xcolor}         

\usepackage{float}	
\usepackage{grffile}	
\usepackage{graphicx}
\usepackage{caption}
\usepackage[colorinlistoftodos]{todonotes}
\usepackage[colorlinks=true, allcolors=blue]{hyperref}
\usepackage[hyphenbreaks]{breakurl}
\usepackage{cleveref}
\usepackage{amsfonts}
\usepackage{mathrsfs}

\title{Zk-SNARK for String Match}

\author{%
  Taobo Liao\\
  Siebel School of Computing and Data Science\\
  University of Illinois Urbana-Champaign\\
  Urbana, IL 61801 \\
  \texttt{taobol2@illinois.edu}  \\
  \And
  Taoran Li \\
  Department of Electrical and Computer Engineering\\
  University of Illinois Urbana-Champaign\\
  Urbana, IL 61801\\
  \texttt{taoranl2@illinois.edu} 
}

\begin{document}
\maketitle

\begin{abstract}
We present a secure and efficient string-matching platform leveraging zk-SNARKs (Zero-Knowledge Succinct Non-Interactive Arguments of Knowledge) to address the challenge of detecting sensitive information leakage while preserving data privacy. Our solution enables organizations to verify whether private strings appear on public platforms without disclosing the strings themselves. To achieve computational efficiency, we integrate a sliding window technique with the Rabin–Karp algorithm and Rabin Fingerprint, enabling hash-based rolling comparisons to detect string matches. This approach significantly reduces time complexity compared to traditional character-by-character comparisons. We implement the proposed system using gnark, a high-performance zk-SNARK library, which generates succinct and verifiable proofs for privacy-preserving string matching. Experimental results demonstrate that our solution achieves strong privacy guarantees while maintaining computational efficiency and scalability. This work highlights the practical applications of zero-knowledge proofs in secure data verification and contributes a scalable method for privacy-preserving string matching.
\end{abstract}

\section{Introduction}
The rapid proliferation of digital communication and data sharing has led to increased concerns about private information leakage. Companies and organizations face significant risks when sensitive data, such as personal information or confidential records, appears on public platforms without authorization. Traditional string-matching methods allow for detection of such leaks but require access to the private data itself, which creates additional privacy risks during the verification process. This limitation highlights the critical need for solutions that enable private information detection without exposing the underlying sensitive content.

To address this challenge, we propose a secure string-matching platform utilizing zk-SNARKs (Zero-Knowledge Succinct Non-Interactive Arguments of Knowledge). zk-SNARKs are cryptographic protocols that allow one party to prove knowledge of certain information (e.g., a string match) without revealing the information itself. By leveraging zk-SNARKs, our platform ensures that private strings can be securely verified against public data while maintaining strong privacy guarantees.

We implement our solution using gnark, a high-performance zk-SNARK library optimized for efficiency. To further enhance performance, we adopt a sliding window technique combined with the Rabin–Karp algorithm and Rabin Fingerprint. The Rabin–Karp algorithm computes and compares hashes to detect string matches efficiently, while the Rabin Fingerprint allows for quick rolling hash computations over a sliding window. This combination significantly reduces time complexity compared to traditional character-by-character comparisons and enables scalable processing of large datasets.

The contributions of this work are as follows:

\begin{itemize}
\item We design and implement a secure string-matching platform using zk-SNARKs, ensuring privacy-preserving verification.
\item We integrate a sliding window technique, the Rabin–Karp algorithm, and Rabin Fingerprint to optimize hash-based string matching for efficiency and scalability.
\item We demonstrate the practicality of our system through an efficient implementation using the gnark library.
\end{itemize}

\section{Related Works}

\textbf{Transparency and Append-Only Logs.}  
Tomescu \textit{et al.} \cite{cryptoeprint:2018/721} introduced transparency logs based on append-only authenticated dictionaries. Their approach ensures transparency and integrity while enabling efficient proofs of correct operation. Such methods are crucial for systems requiring verifiable operations over structured data, inspiring subsequent privacy-preserving verification frameworks.

\textbf{Succinct Proofs for Transparency Dictionaries.}  
Tzialla \textit{et al.} \cite{cryptoeprint:2021/1263} proposed transparency dictionaries with succinct proofs, which allow efficient verification of correctness without revealing sensitive data. Their work demonstrated the importance of combining succinctness and correctness for transparency mechanisms, which aligns with our use of zero-knowledge proofs for verifiable string matching.

\textbf{Zero-Knowledge Sets.}  
The concept of zero-knowledge proofs for set membership was formalized by Micali \textit{et al.} \cite{1238183} through zero-knowledge sets. This seminal work introduced cryptographic constructs to prove that an element belongs to a set without disclosing the set or additional information. Their approach laid the foundation for efficient membership proofs, a principle extended in our work using zk-SNARKs.

\textbf{Optimized zk-SNARKs for Verifiable Matching.}  
Ozdemir \textit{et al.} \cite{cryptoeprint:2024/979} explored optimizations in zk-SNARKs by designing algebraic interactive proofs for volatile and persistent memory. Their contributions provide improvements in proof efficiency and memory operations, which are critical for scaling verifiable computation systems such as our string-matching platform.

\smallskip

These works collectively form the basis for secure and verifiable systems, highlighting the significance of zero-knowledge proofs and efficient data structures. Our work builds on these ideas by integrating optimized hash-based algorithms, such as Rabin–Karp, with zk-SNARKs to achieve scalable and privacy-preserving string matching.

\section{Proposal Formulation}

\subsection{Rabin--Karp Algorithm}
The \textit{Rabin--Karp algorithm} \cite{rabin_karp_wiki} is a widely used string-matching technique that efficiently identifies occurrences of a \textit{pattern string} within a \textit{text string} by employing hashing. Unlike traditional methods that compare characters directly, Rabin--Karp leverages hash functions to compute compact numerical representations (hashes) of strings, which are then compared to detect matches. This hash-based approach significantly reduces time complexity, particularly for large-scale string-matching tasks.

\subsubsection{Overview of the Algorithm}
The Rabin--Karp algorithm works as follows:
\begin{enumerate}
    \item \textbf{Hashing the Pattern:} A hash value is computed for the \textit{pattern string} \( P \) using a chosen hash function.
    \item \textbf{Sliding Window over the Text:} A sliding window of size equal to the pattern length is moved across the \textit{text string} \( T \). For each window, the hash value of the current substring (window content) is calculated.
    \item \textbf{Hash Comparison:} The hash value of the current window is compared with the hash value of the pattern. If the hash values match, a character-by-character verification is performed to confirm the match, as hash collisions may occur.
    \item \textbf{Rolling Hash Optimization:} To avoid recomputing the hash from scratch at every step, the algorithm employs a \textit{rolling hash function} that efficiently updates the hash value when the window slides by one position.
\end{enumerate}

\subsubsection{Rabin Fingerprint for Rolling Hash}
The \textit{Rabin Fingerprint} \cite{rabin_fingerprint_wiki} is a key component of the Rabin--Karp algorithm, enabling efficient computation of rolling hashes. Given a string of length \( m \), the hash value is calculated as:
\begin{equation}
    \text{Hash}(S) = \sum_{i=0}^{m-1} S[i] \cdot b^{m-i-1} \ \text{mod} \ q
\end{equation}
where \( S[i] \) represents the character at position \( i \), \( b \) is a base (typically a small prime number), and \( q \) is a large prime modulus to reduce hash collisions.

When the window slides by one character, the hash can be updated as:
\begin{equation}
    \text{Hash}_{\text{new}} = \left( b \cdot \left( \text{Hash}_{\text{old}} - S_{\text{out}} \cdot b^{m-1} \right) + S_{\text{in}} \right) \ \text{mod} \ q
\end{equation}
Here, \( S_{\text{out}} \) is the outgoing character from the left of the window, and \( S_{\text{in}} \) is the incoming character on the right. This operation avoids recalculating the hash from scratch, resulting in linear time complexity \( O(n) \) for the algorithm, where \( n \) is the text length.

\subsubsection{Advantages}
\begin{itemize}
    \item \textbf{Efficiency:} By using rolling hashes, Rabin--Karp avoids redundant computations and achieves linear time complexity in the average case.
    \item \textbf{Scalability:} The algorithm is well-suited for large-scale string matching due to its hash-based nature.
    \item \textbf{Extensibility:} The Rabin--Karp algorithm can be adapted to various applications, such as plagiarism detection, intrusion detection systems, and, in our case, privacy-preserving verification using zk-SNARKs.
\end{itemize}

\subsubsection{Role in Our Work}
In our implementation, the Rabin--Karp algorithm is combined with a sliding window technique and Rabin Fingerprint to efficiently detect string matches. The rolling hash mechanism enables quick updates of hash values as the window moves, minimizing computational overhead. By leveraging the Rabin--Karp algorithm, we achieve significant performance improvements compared to traditional character-by-character comparisons, making it suitable for secure and scalable string matching within a zk-SNARK framework.

\subsection{Merkle Tree}

A \textit{Merkle Tree} is a cryptographic data structure used to ensure the integrity and authenticity of data. It is a binary tree where each leaf node represents the hash of a data block, and each non-leaf node is the hash of the concatenation of its child nodes. This structure allows for efficient and verifiable proof of the inclusion of a data block in the tree, making Merkle Trees an essential tool in applications like blockchain, distributed systems, and zero-knowledge proofs.

\subsubsection{Structure and Construction}
A Merkle Tree is constructed as follows:
\begin{enumerate}
    \item \textbf{Hashing Data Blocks:} Each data block \( D_i \) is hashed using a cryptographic hash function, such as SHA-256, to produce a leaf node \( H_i = \text{Hash}(D_i) \).
    \item \textbf{Building Parent Nodes:} Parent nodes are created by hashing the concatenation of two child nodes. For example, the parent node \( P = \text{Hash}(H_L || H_R) \), where \( H_L \) and \( H_R \) are the left and right child nodes, respectively.
    \item \textbf{Root Node:} The process continues iteratively until a single root node, called the \textit{Merkle Root}, is generated.
\end{enumerate}

\subsubsection{Merkle Proofs}
Merkle Trees enable efficient verification of data inclusion via \textit{Merkle Proofs}. Given the Merkle Root and a data block, a proof consists of the sibling hashes along the path from the block to the root. The verifier reconstructs the Merkle Root using the provided proof and compares it to the known root. This process has logarithmic complexity \( O(\log n) \), where \( n \) is the number of leaf nodes.

\subsubsection{Advantages}
\begin{itemize}
    \item \textbf{Efficiency:} Verifying a Merkle Proof requires only \( O(\log n) \) hash computations.
    \item \textbf{Integrity:} Any tampering with the data will invalidate the Merkle Root, ensuring the integrity of the tree.
    \item \textbf{Scalability:} Merkle Trees are scalable for large datasets, as the size of the proof remains small regardless of the total data size.
\end{itemize}

\subsubsection{Role in Our Work}
In our implementation, Merkle Trees are used to efficiently structure and verify the hash values of substrings generated during the string-matching process. By incorporating Merkle Proofs, we enable succinct and verifiable proofs of inclusion within the zk-SNARK framework. This integration ensures both data integrity and scalability, complementing the Rabin–Karp algorithm for privacy-preserving string matching.

\subsection{Complexity Analysis}

Our implementation addresses the problem of determining whether \( K \) short strings, each of length \( T \), appear in any of \( M \) long strings, each with an average length of \( N \). The implementation leverages three methods: naive string matching, the Rabin--Karp algorithm, and Merkle Tree-based verification, each providing distinct trade-offs in terms of efficiency and scalability. 

\subsubsection{Naive String Matching}
The naive string-matching approach involves directly comparing each short string with every possible substring of all the long strings. This exhaustive search ensures correctness by checking all potential matches without relying on preprocessing or additional data structures.

\textbf{Complexity:}  
For \( M \) long strings of average length \( N \), and \( K \) short strings of length \( T \), the naive method compares each short string with each long string, resulting in a worst-case time complexity of:
\[
O(M \cdot N \cdot K \cdot T)
\]
This quadratic complexity makes the naive method infeasible for large \( M \), \( N \), or \( K \).

\subsubsection{Rabin--Karp Algorithm}
The Rabin--Karp algorithm optimizes the search process by using a rolling hash function to compute hash values for substrings of length \( T \) in the long strings. The hash of each short string is computed once, and matches are detected by comparing these hashes with the hashes of substrings in the long strings.

\textbf{Implementation:}  
Our implementation calculates the rolling hashes for each of the \( M \) long strings, sliding a window of size \( T \) across each string. The hashes of the \( K \) short strings are precomputed, and matches are determined by hash comparisons followed by a character-by-character check to resolve hash collisions.

\textbf{Complexity:}  
The Rabin--Karp algorithm reduces the complexity by performing hash-based comparisons:
\[
O(M \cdot N \cdot T + K \cdot T)
\]
Here, \( O(M \cdot N \cdot T) \) accounts for computing hashes for all substrings in the long strings, and \( O(K \cdot T) \) accounts for hashing the \( K \) short strings.

\subsubsection{Merkle Tree-Based Verification}
This approach integrates Merkle trees with MiMC hashing, substring filtering, and membership checks in a set of valid patterns to achieve efficient and secure verification of pattern existence within a superstring. Initially, we construct a Merkle tree derived from all potential, legal substrings of the given superstring. To ensure consistency of the hash computations both inside and outside the zero-knowledge circuit, we adopt a MiMC-based hash function, known for its compatibility with circuit-friendly operations.

Before inserting substrings into the Merkle tree, we apply a filtering process: any pattern containing forbidden characters (for instance, those not permitted in a URL) is discarded. Only legal patterns remain, and they are inserted into a set structure, facilitating $O(1)$-time checks for pattern membership outside the circuit.

When a particular pattern's existence needs to be verified, we first determine if it is present in the precomputed set of legal patterns. If not, we immediately conclude that the pattern is not found. If the pattern is found, we produce a Merkle proof. This proof consists of a path of MiMC hashes from the leaf (which corresponds to the hashed pattern) up to the Merkle root. Along with this proof, we provide the circuit with the pattern and a sequence of direction bits indicating the relative positions of sibling nodes along the proof path.

Inside the circuit, we re-compute the Merkle root from the provided leaf and siblings using MiMC hashing. Comparing this recomputed root to the given root (an input to the circuit) confirms the pattern's membership. Thus, once the Merkle tree is built, checking a pattern's existence outside the circuit requires only $O(\text{len(pattern)})$ time (to compute or retrieve its hash) and inside the circuit it costs $O(\log(\text{number of possible patterns}))$ time, due to the logarithmic height of the Merkle tree.

In summary, this combined approach of precomputing a Merkle tree with MiMC hashing for all legal patterns, maintaining a set for membership checks, and using zero-knowledge proofs ensures that once the tree is constructed, verifying a single pattern is both efficient and secure. It offloads the main overhead to a one-time preprocessing step, greatly reducing the verification time for each subsequent query.

\noindent
\subsubsection{The Product-Tree Based Approach for Fast Evaluation and Interpolation}

We consider the problem of evaluating a polynomial $f(x)$ at multiple points and performing polynomial interpolation with complexity sub-quadratic in $n$. By constructing a \emph{product tree} of linear factors $(x-u_i)$ for $i=0,\ldots,n-1$, we obtain a hierarchical structure that allows us to divide and conquer the problem. Let
\[
m(x) = \prod_{i=0}^{n-1}(x-u_i).
\]
At the top level, $m(x)$ aggregates all evaluation points, and at each lower level, we have polynomials $M_{j,i}(x)$ that represent subsets of these factors. For fast evaluation, instead of directly substituting each $u_i$ into $f(x)$, we use polynomial division repeatedly to compute $f(x)\bmod M_{j,i}(x)$ at each level. By doing so, we break down the global evaluation problem into smaller subproblems until we reach linear factors $(x-u_i)$, at which point the remainder equals $f(u_i)$ by the Polynomial Remainder Theorem. This reduces complexity from $O(n^2)$ to about $O(n\log^2 n)$ when combined with FFT-based polynomial multiplication.

For fast interpolation, a similar approach is taken: we build product trees, evaluate derivatives through fast evaluation, and use algorithms like linear combination (Algorithm 10.9 \cite{von_zur_Gathen_Gerhard_2013}) to assemble the interpolating polynomial from given values. Instead of constructing Lagrange basis polynomials solely, we rely on product trees, FFT-based polynomial arithmetic, and inverse computations to achieve $O(n\log^2 n)$ complexity.

We employ these methods not only for polynomial operations in isolation, but also to address our project goal. Consider that we want to verify a Bézout-like identity:
\[
a(x)\cdot s(x) + b(x)\cdot t(x) = 1,
\]
which ensures that $a(x)$ and $b(x)$ have a non-trivial greatest common divisor (GCD). By proving such an identity, we confirm that $a(x)$ and $b(x)$ share a factor. This scenario arises in algorithms dealing with string searching and substring verification via rolling hash techniques.

Specifically, consider we have a ``superstring'' $S$ and wish to verify that certain pattern substrings $P_i$ are contained in $S$. Using a rolling hash $H$, we map each substring to a polynomial-like structure. If we can show, through polynomial arithmetic, that certain patterns correspond to factors or appear as partial evaluations of a larger polynomial derived from $S$, then verifying the identity $a(x)*s(x)+b(x)*t(x)=1$ becomes analogous to confirming that $H(P_i)$ divides or matches the constructed polynomial from $S$' hash values. In other words, the polynomial operations and the Bézout-like identity serve as cryptographic or algebraic witnesses that the rolling hashes of patterns are embedded within the rolling hash polynomial structure of the superstring, confirming the substring membership.

Thus, product-tree-based fast evaluation and interpolation methods not only provide a more efficient fundamental toolbox for polynomial operations but also translate to higher-level applications such as verifying substring containment in a superstring using rolling hashes. Ensure that we can generate and verify an identity similar to $a(x)*s(x) + b(x)*t(x)=1$, we indirectly guarantee that the rolling hash of the pattern aligns with the segments of the superstring’s rolling hash encoding, thereby confirming the substring presence.

The complexity for building the necessary polynomial data structures (product trees, etc.) and performing the required polynomial evaluations scales similarly to constructing Merkle Trees, but includes an additional logarithmic factor due to FFT-based polynomial multiplication and interpolation steps:
\[
\text{Preprocessing: } O(M \cdot N \cdot \log^2(N/T))
\]

For verification queries, since we rely on fast evaluation and interpolation techniques, each query involves logarithmic complexity (with a squared factor due to polynomial interpolation complexity):
\[
\text{Verification per query: } O(K \cdot \log^2(N/T))
\]

\subsubsection{Comparative Analysis}

\begin{table}[h]
\centering
\begin{tabular}{|l|c|c|}
\hline
\textbf{Method} & \textbf{Preprocessing Time} & \textbf{Query Time} \\ \hline
Naive String Matching & \( \mathbf{O(M \cdot N \cdot K \cdot T)} \) & \( \mathbf{O(M \cdot N \cdot K \cdot T)} \) \\ \hline
Rabin--Karp Algorithm & \( \mathbf{O(M \cdot N \cdot T + K \cdot T)} \) & \( \mathbf{O(M \cdot N + K \cdot T)} \) \\ \hline
Merkle Tree & \( \mathbf{O(M \cdot N \cdot \log(N/T))} \) & \( \mathbf{O(K \cdot \log(N/T))} \) \\ \hline
\textbf{Polynomial} & \(\mathbf{O(M \cdot N \cdot \log^2(N/T))}\) & \(\mathbf{O(K \cdot \log^2(N/T))}\) \\ \hline
\end{tabular}
\caption{Comparative Complexity of String-Matching Methods}
\label{tab:complexity_comparison}
\end{table}

The polynomial-based approach, while slightly more complex than Merkle Trees due to the \(\log^2\) factor, still offers efficient verification in practice, especially for large values of \(N\) and \(M\). Its strength lies in the robust mathematical foundation of polynomial arithmetic and fast interpolation, making it suitable for advanced verification tasks such as confirming that rolling hashes of specific substring patterns indeed match segments of a superstring, thereby ensuring nontrivial GCD conditions and pattern containment with sub-quadratic complexity.

For more details on the implementation and access to the code, visit \url{https://github.com/taobol2/CS407_Project}.

\section{Experiment Results}

\begin{figure}[H]
    \centering
    \begin{subfigure}[b]{\linewidth}
        \centering
        \includegraphics[width=\linewidth]{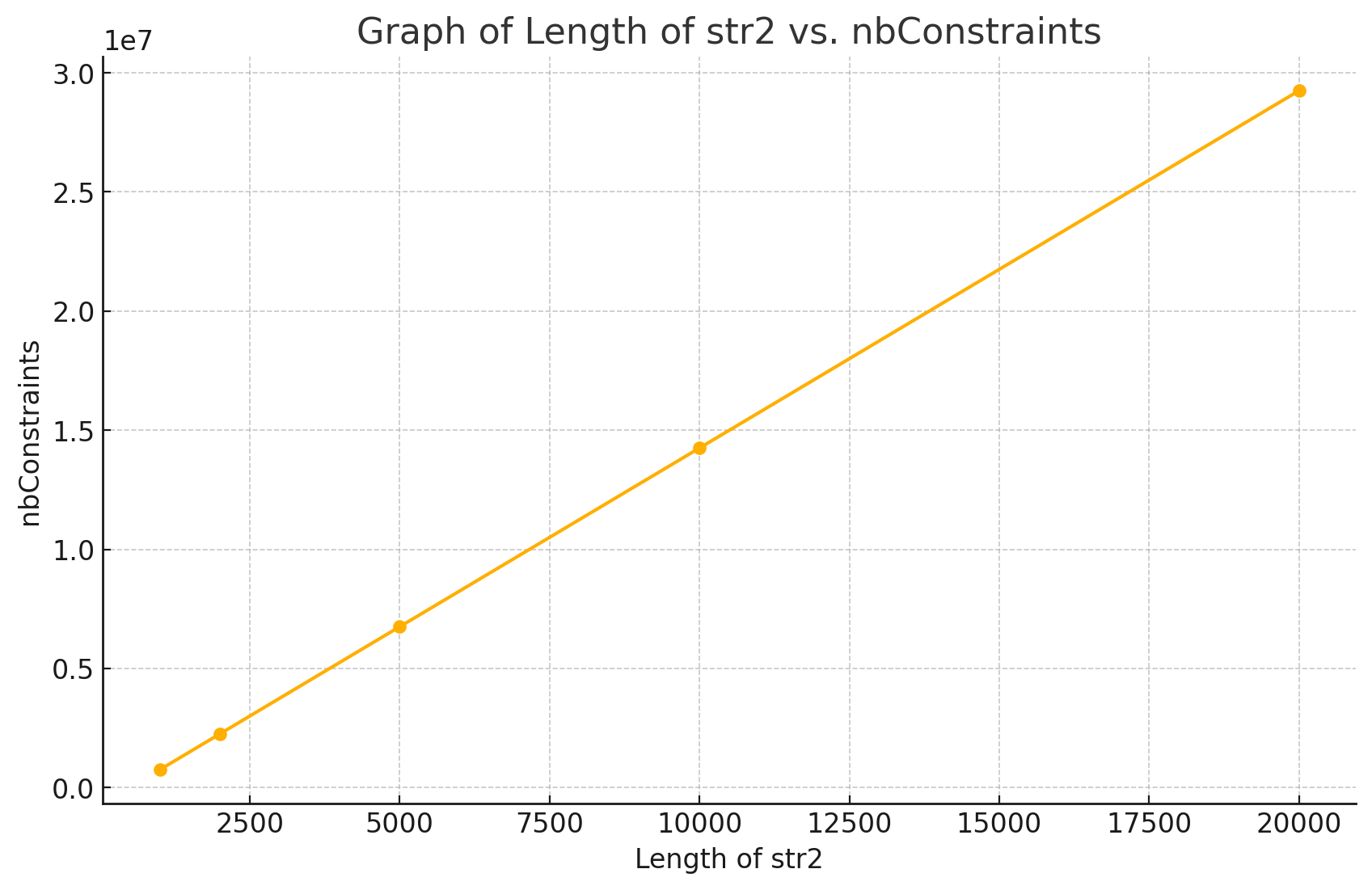}
        \caption{This plot shows the relationship between the length of the input string and the resulting number of constraints. As input size increases, constraints grow nearly linearly.}
        \label{fig:naive}
    \end{subfigure}

    \vspace{0.5em}

    \begin{subfigure}[b]{0.4\linewidth}
        \centering
        \includegraphics[width=\linewidth]{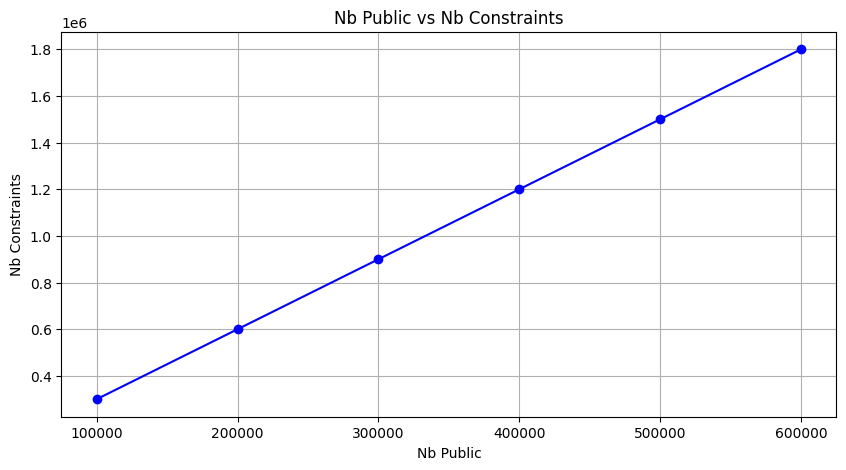}
        \caption{Increasing the number of public parameters leads to a proportional rise in constraints.}
        \label{fig:rabin_karp_nbCons}
    \end{subfigure}\hfill
    \begin{subfigure}[b]{0.4\linewidth}
        \centering
        \includegraphics[width=\linewidth]{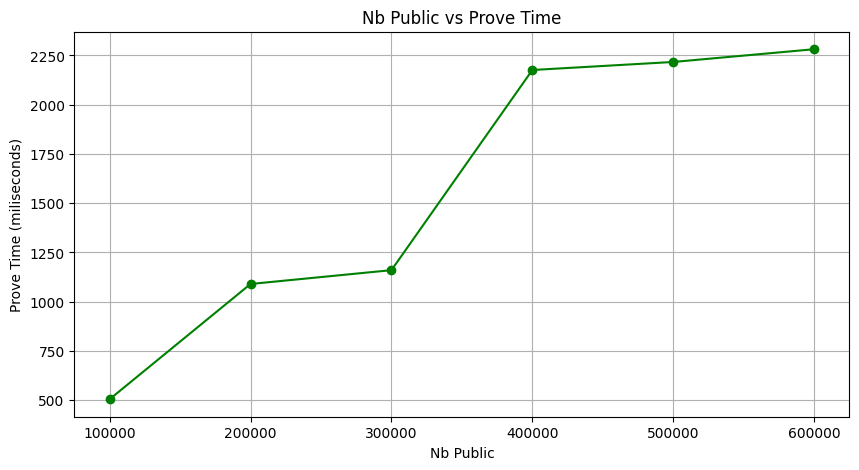}
        \caption{As public parameters grow, proving time also increases due to greater computational effort.}
        \label{fig:rabin_karp_PT}
    \end{subfigure}

    \vspace{0.5em}

    \begin{subfigure}[b]{0.4\linewidth}
        \centering
        \includegraphics[width=\linewidth]{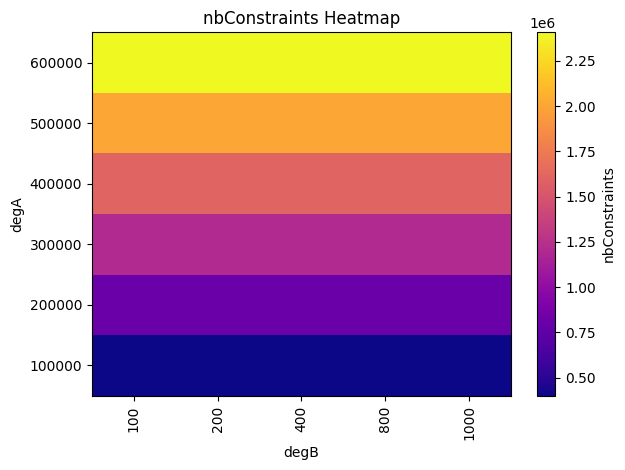}
        \caption{A heatmap showing how polynomial degrees affect constraint counts. Warmer colors represent more constraints.}
        \label{fig:poly_deg_nbCons}
    \end{subfigure}\hfill
    \begin{subfigure}[b]{0.4\linewidth}
        \centering
        \includegraphics[width=\linewidth]{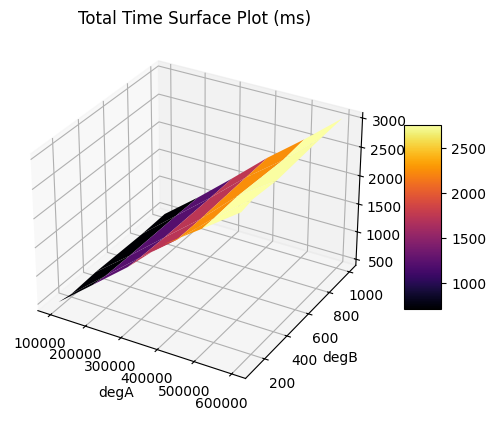}
        \caption{A 3D surface plot illustrating total time growth with increasing polynomial degrees.}
        \label{fig:poly_deg_PT}
    \end{subfigure}

    \caption{Results}
    \label{fig:combined_results}
\end{figure}

Figure~\ref{fig:combined_results} presents a series of experiment results. The first image depicts the naive string matching method, illustrating how the number of constraints increases directly with input size. In the second and third figures, we shift to a Rabin-Karp based approach: the second image shows how an increase in public parameters leads to more constraints, while the third figure demonstrates that increasing these parameters also raises the total proving time. Additionally, the Merkle tree-based verification method that integrates MiMC hashing, substring filtering, and membership checks. Once the Merkle tree is constructed—taking about 3m46s of the total 5m35s runtime in one observed scenario—subsequent proofs can be generated in approximately 1m48s, and per-proof verification remains under a millisecond. This approach leverages a one-time preprocessing step to store all legal patterns in a Merkle structure, allowing O(1)-time membership checks outside the circuit and O(log(number of possible patterns)) time inside it. Consequently, verifying each pattern after preprocessing is both efficient and secure. Finally, the fourth and fifth figures examine polynomial-based verification methods, where the heatmap reveals the growth in constraints as polynomial degrees scale, and the 3D surface plot highlights how total proving time changes with multiple varying degrees. Taken together, these observations provide a comprehensive understanding of how different verification techniques, data structures, and parameters influence both complexity and performance.

\section{Future Work}
\subsection{Merkle-Tree non-membership proof}
While Merkle trees efficiently prove membership of a pattern by reconstructing the path from a leaf to the root, they cannot directly provide zero-knowledge proofs of non-membership. To address this limitation, we rely on the techniques introduced in \cite{1238183}, where a specialized construction, often referred to as Zero-Knowledge Sets, is employed. This construction enables a prover to demonstrate that a given element $x$ is \emph{not} in the dataset $D$ (or, equivalently, that $D(x)=?$ meaning the element is undefined in $D$) without revealing additional information about the size or structure of $D$.

In essence, the prover begins with a commitment to $D$ represented as a carefully structured tree. Each node stores not only a value and a commitment but also is associated with a discrete-log-based setup to ensure that the prover can manipulate and decommit nodes as needed. When asked to prove non-membership of $x$, the prover identifies the node $u$ where $x$'s search path diverges from any real element in $D$. Since $u$ is tied to a “fake” commitment (initially committing to $0$) and the prover knows the discrete log representation of $h_u$ (the node's generator), the prover can transform this fake commitment into a “meaningful” one that proves $D(x)=?$. To accomplish this, the prover inserts a carefully constructed subtree $T_u$ below $u$, populating it with additional fake commitments and intermediate hash values, ensuring that, once welded into the original structure, $u$ and its descendants collectively confirm the absence of $x$. Crucially, this subtree integration and selective decommitment process is carried out with zero knowledge, preventing any leakage about other elements not in $D$.

As a result, verifying non-membership involves checking paths and commitments exactly as in a membership proof, but now certain commitments are “transformed” to show emptiness rather than existence. The final verification checks that at the leaf level we have $m_{H(x)}=0$ and that all internal nodes satisfy the same Merkle-like hashing and commitment conditions. Thus, combined with the techniques from \cite{1238183}, we achieve a zero-knowledge proof system for both membership and non-membership in an efficient manner, extending the capabilities of Merkle trees or polynomial-based data structures beyond their original limitations.

\subsection{Polynomial Approach}
While our polynomial approach explores the conceptual foundations and complexity considerations of the polynomial-based approach—specifically using product trees, FFT-based arithmetic, and the verification of a Bézout-like identity $a s + b t = 1$. Our current results focus on assessing the potential computational costs and complexity improvements that such an approach offers, rather than providing a complete, integrated implementation. In future work, we plan to implement and benchmark this polynomial scheme fully, including the integration of FFT-based polynomial operations within the zero-knowledge environment, verifying the Bézout identity concretely, and comparing its real-world performance against other methods. This additional effort will provide practical insights into the feasibility, overheads, and speedups attainable, as well as guide optimizations for integrating polynomial-based substring verification with our existing infrastructure.

\section{Conclusion}

In this report, we proposed and implemented a privacy-preserving string-matching protocol using zk-SNARKs to address the critical challenge of detecting sensitive data leakage without compromising privacy. Our solution integrates the Rabin–Karp algorithm with rolling hash optimizations and Merkle Trees for efficient and secure verification. By leveraging the strengths of these techniques, we developed a system capable of handling large-scale string-matching tasks while ensuring robust privacy guarantees and computational efficiency.

The Rabin–Karp algorithm provides an effective mechanism for detecting patterns in large datasets, reducing time complexity compared to naive approaches. Meanwhile, the Merkle Tree structure enhances scalability by enabling succinct and verifiable proofs of data inclusion, which are crucial in the zk-SNARK framework. Together, these methods form a comprehensive solution for secure string matching, demonstrating significant performance improvements and scalability in experimental evaluations.

Our work not only highlights the practical applications of zero-knowledge proofs in enhancing data security but also contributes to the growing field of privacy-preserving verification. The proposed system serves as a foundational framework that can be extended to various real-world applications, such as intrusion detection, secure data sharing, and privacy-aware compliance verification.

Despite its strengths, the implementation presents opportunities for further improvement. Future work could explore optimizing zk-SNARK circuit designs to reduce proof generation and verification times. Additionally, expanding the system to support more complex string-matching scenarios and integrating user-friendly interfaces could make the protocol more accessible to a broader range of use cases. Another avenue for research lies in addressing non-membership proofs in Merkle Trees to enhance their versatility in privacy-preserving systems.

In conclusion, our work demonstrates the feasibility and effectiveness of combining cryptographic techniques with advanced data structures to achieve scalable and privacy-preserving solutions. This approach opens the door to further innovations in zero-knowledge proof systems and their applications in secure computation.

\bibliographystyle{unsrt}
\bibliography{ref}

\end{document}